\documentclass[fleqn]{aa} 
\usepackage{epsf}
\usepackage{graphicx}

\newcommand{\msun}{$M_{\odot}$}

\newcommand{\prot}{$P_{\rm rot}$}

\newcommand{\teff}{$T_{\rm eff}$}

\newcommand{\vrot}{$v_{\rm rot}$}

\newcommand{\vr}{$V_{\rm R}$}
\newcommand{\ir}{$I_{\rm R}$}

\newcommand{\lp}{\mbox{LP\,790-29}}

\begin{document}
\title{The high-field magnetic white dwarf LP790-29:\\ not a fast
rotator \thanks{Based on observations collected at the European
Southern Observatory, Chile, under programme ID 64.H--0311.}  }
\author{K. Beuermann
\and K. Reinsch
} 
\offprints{beuermann@uni-sw.gwdg.de}
\institute{ 
Universit\"ats-Sternwarte, Geismarlandstr. 11, D-37083 G\"ottingen, Germany
}
\date{Received September 21, 2001 / Accepted October 22, 2001}
\authorrunning{K. Beuermann \& K. Reinsch}
\titlerunning{LP790-29, not a fast rotator}
\abstract{We have investigated the nature of the magnetic white dwarf
\lp\ = LHS\,2293 by polarimetric monitoring, searching for short-term
variability. No periodicity was found and we can exclude rotation
periods between 4 sec and 1.5\,hour with a high confidence. Maximum
amplitudes of sinusoidal variations are $\Delta R < 0.009$\,mag and
$\Delta V_{\rm R} < 0.7$\,\% for a mean value of the $R-$band circular
polarization of $V_{\rm R} = +9.1\pm0.3$\,\%. Combined with earlier
results by other authors, our observation suggests that \lp{} is, in
fact, an extremely slowly rotating single white dwarf and not an
unrecognized fast rotator and/or disguised cataclysmic variable.
}

\maketitle

\section{Introduction}

The large majority of white dwarfs are slow rotators with equatorial
velocities \vrot\,sin\,i~$<\,15$\,km\,s$^{-1}$ and rotational periods
\prot\,$ \ga 1$\,hr (Heber et al. 1997, Koester et al. 1998). Much slower
rotation is not detectable in non-magnetic white dwarfs, but easily
measureable in magnetic ones by polarimetric monitoring (Schmidt \&
Norsworthy 1991, Berdyugin \& Piirola 1999).

Among the magnetic white dwarfs, there is a surprising dichotomy in
the distribution of \prot\ for magnetic white dwarfs with all stars
having either \prot\,$<\,20$\,d or \prot\,$>\,100$\,yrs (Schmidt \&
Norsworthy 1991). Some magnetic white dwarfs rotate surprisingly fast
while others are apparently extremely slow. Among the fast ones are
the DA white dwarf RE0317-853 with $B \simeq 500$ MG, \teff\, $\simeq
40\,000$\,K, and \prot\,= 12 min (Barstow et al. 1995, Ferrario et
al. 1997, Burleigh et al. 1999), the DA star PG1015+014 with $B \simeq
120$ MG, \teff\,$\simeq 14\,000$\,K, and \prot\,= 99 min
(Wickramasinge \& Cropper 1988, Schmidt and Norsworthy 1991), and the
DAB white dwarf Feige\,7 with $B \sim 35$ MG, \teff\,$\simeq
20\,000$\,K, \prot\,= 2.2\,hr (Liebert et al. 1977, Achilleos et
al. 1992). Five systems seem to be very slow rotators with \prot\,$ >
100$\,yr (Schmidt \& Norsworthy 1991), among them the proven systems,
GD229, G240-72 (Berdyugin \& Piirola 1999), and Grw+$70^{\circ}8247$
(Friedrich \& Jordan 2001), as well as a suspected one, LP790-29
(Liebert et al. 1978).

Slow rotation may be caused by coupling of angular momentum into the
giant envelope of the progenitor star or the interstellar medium
during later stages (Schmidt \& Norsworthy 1991). Fast rotation may be
achieved in a double degenerate which ends as a merger (e.g. Segretain
et al. 1997) or in a magnetic cataclysmic variable which loses
synchronism (Meyer \& Meyer-Hofmeister 1999). If the donor in a
mass-transfer binary is nearly substellar and hydrogen rich
(short-period AM Herculis binary), the white dwarf is expected to be
of spectral type DA which is not the case for LP790-29.
Cataclysmic variables with a (partially) degenerate low-mass donor (AM
CVn binaries) transfer helium or carbon and typically end as CO white
dwarfs, possibly with a substellar companion (Iben \& Tutukov
1991). The hot white dwarf RE0317-853 has been suggested to be the
result of a merger (Barstow et al. 1995, Ferrario et al. 1997) or a
mass-transfer binary (Meyer \& Meyer-Hofmeister 1999). The former
appears more likely because the primary in RE0317-853 is hot with
\teff\,$\simeq 40\,000$ K and all white dwarfs in short-period
cataclysmic variables are cool with \teff\,$\simeq 9\,000 -
15\,000$\,K (G\"ansicke 2000). Although definite conclusions in any
individual case may be problematic, the detection of rapid rotation of
magnetic white dwarfs would clearly help to elucidate their
evolutionary history. It also helps to understand the physical
processes by which angular momentum is coupled into the environment of
the star (Schmidt \& Norsworthy 1991) and may help to define sources
of gravitational wave radiation (Heyl 2000).

\begin{figure*}[t]
\begin{center}
\includegraphics[width=12cm]{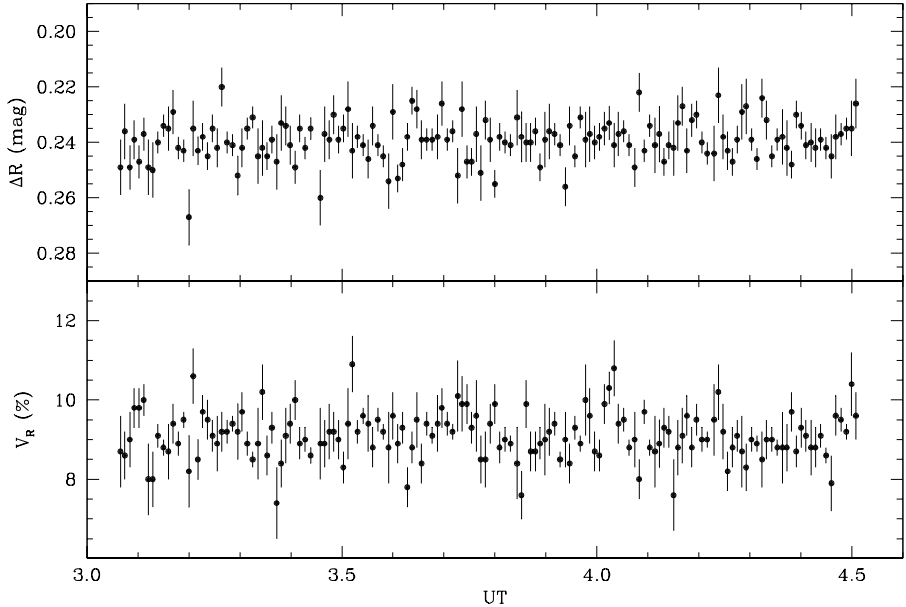}
\caption{Time series of the Stokes intensity $I_{\rm R}$ and circular
polarization $V_{\rm R}$ in the Bessell $R$-band. The time resolution
is \mbox{$\sim 30-40$\,sec} with 150 individual CCD images taken over
1.5\,h. The $R$-band magnitude was measured relative to two comparison
stars in the field of \lp.}
\label{lightcurves}
\end{center}
\end{figure*}

\lp\,=\,LHS\,2293 was discovered by Liebert et al. (1978) and found to
be a highly circularly polarized cool white dwarf which shows the
Zeeman shifted C$_2$ Swan bands. It has Stokes $V\,\simeq\,+8$\% and
+9\% at wavelengths shortward of 4300\,\AA\ and longward of 5500\AA,
respectively, (Liebert et al. 1978, West 1989) which decreases to nil
inbetween and to 2\% in the J--band. Bues (1999) refined its
temperature to \teff \,$\simeq 7800$\,K. The field strength was
originally quoted as $B \sim 200$\,MG (Liebert et al. 1978, Schmidt \&
Smith 1995), while Bues used 50\,MG in her spectral fitting, and
Wickramasinghe \& Ferrario (2000) quoted an uncertain 100\,MG. \lp\ is
also linearly polarized at the level of $\sim 1$\,\% (West
1989). Polarimetric observations (Liebert et al. 1978, Robert \&
Moffat 1989, West 1989) have shown that the level of circular
polarization has stayed constant over 10 years, excluding rotation
periods in the range of $\sim $20\,min\,$\la P_{\rm rot}\,\la\,100$\,yrs
(Schmidt \& Norsworthy 1991). In a non-axisymmetric field
geometry, the circular polarization will depend on rotation phase and
a short rotation period should be readily detectable in time series of
the circular polarization, provided the magnetic axis is inclined for
more than a few degrees against the rotational axis and the latter
does not point directly at the observer. 
In this communication, we report results of a search for rapid
rotation in \lp, using photometric and polarimetric data taken in the
Bessell $R-$band. Given the observed spectral dependence of
Stokes $V$ (Liebert et al. 1978, Schmidt et al. 1995), the $R-$band
provides the best polarimetric signal of the standard photometric
bands.

\section{Observations and Results}

We observed \lp{} on February  4, 2000, with the ESO 3.6m telescope
equipped with the focal-reducer spectrograph and camera, EFOSC2, and a
user-supplied superachromatic quarter-wave plate which was produced by
Halle/Berlin and is of the same type as used in the ESO VLT FORS1
spectrograph. Flux and circular polarization were measured in the
photometric Bessell R-band with a time resolution of $\sim$30--40\,sec
for a total of 1.5 h.  Pairs of images with the retarder-plate
position angle alternating between $-45\degr$ and $+45\degr$ were
taken with exposure times of 2\,s, 3\,s, 5\,s, and 10\,s chosen in
random order.  The dead time between two exposures was 28--32\,sec due
to readout and instrument-setup times. Reading out single exposures
guaranteed the best possible $S/N$ ratio for Stokes \vr. The $R-$band
was chosen because of the higher level of circular polarization
compared with that at the shorter wavelengths (Liebert et al. 1978).
The measured values of \vr\ have been corrected for instrumental
biases and linear-polarization cross-talk, $-1.3\pm 0.3$\%, which was
determined to the first order from each pair of observations. The
error in \vr\ is dominated by this systematic uncertainty, its
statistical error is $< 0.1$\,\%.

\begin{figure*}[t]
\begin{center}
\includegraphics[width=15cm]{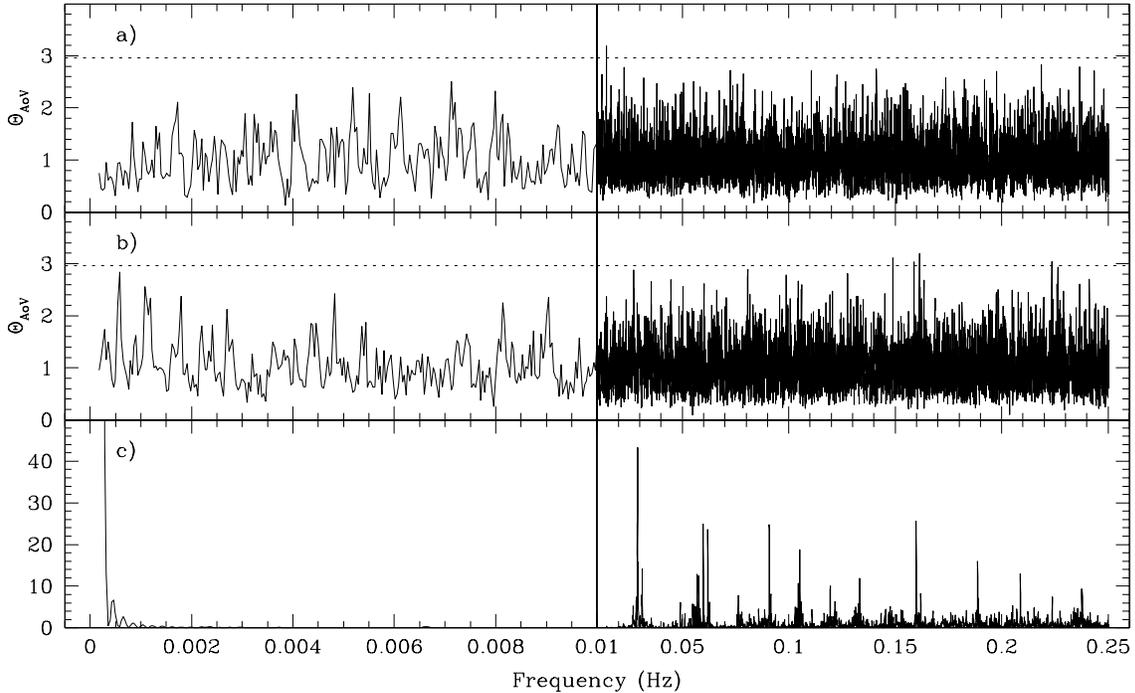}
\caption{Periodogram of (a) the Stokes intensity $I_{\rm R}$ and (b)
the circular polarization $V_{\rm R}$ in the Bessell $R$-band, along
with (c) the spectral window function of the data. The short-dashed
lines in panels a) and b) give the Fisher-Snedecor critical value for
a 3-$\sigma$ detection of a periodic signal in the data, uncorrected
for bandwidth (see text). The right hand panel covers periods from
4\,sec to 100\,sec, the left hand panel provides an expanded view for
periods from 100\,sec to 1.5\,hours}
\label{periodograms}
\end{center}
\end{figure*}

Figure \ref{lightcurves} shows the resulting time series. The $R-$band
magnitude was measured against two comparison stars with similar
brightness, USNO 0675\_11099946 and USNO 0675\_11100282, located 14
arcsec SW and 34 arcsec NE of \lp.  In order to search for periodicities, we
computed the Analysis of Variance statistics implemented in the
European Southern Observatory Munich Image Data Analysis System
(MIDAS) software package (Schwarzenberg-Czerny 1989) for periods
between the Nyquist limit of \prot\,=\,4\,sec and a maximum period
\prot\,=\,1.5\,h (Figure \ref{periodograms}).  The expected value of
this statistic for pure noise is $\Theta_{\rm AoV} = 1$. The critical
value of the Fisher-Snedecor distribution for a 3-$\sigma$ detection
of a periodic signal in the data is $\Theta_{\rm cr} = 2.96$ for 9 and
140 degrees of freedom in the numerator and the denominator,
respectively.  The lack of a slope or curvature in the data excludes
periods up to $\sim\,3$\,h.
%

In order to estimate the upper limit on the amplitude of any
periodicity in the range of 4\,sec to 1.5\,hour, we added artificial
sinusoidal signals with various periods to the data and repeated the
periodogram analysis. Amplitudes in excess of $\Delta R
=$\,0.006--0.009 mag and $\Delta V_{\rm R} =$\,0.5--0.7\,\% can be
excluded, with the lowest sensitivity corresponding to periods close
to the local maxima of the spectral window function.  Hence, we find
that any coherent periodicity present must have a fractional
modulation of Stokes $V_{\rm R} < 0.7$\%.

No significant photometric or polarimetric variability was
detected. The periodogram in Fig. 2 is divided into two sections of
which the left-hand one covers the more relevant longer periods that
might be expected for a merger. The right-hand panel demonstrates
that no periodicity is found down to \prot\ = 4\,sec, the break-up
period of a 1.2\,\msun\ white dwarf. For 6.950 trial periods each, in
the photometric and in the polarimetric data, only a few frequency
bins reach a significance above $3\,\sigma$, entirely consistent with
expectation for the detection of spurious lines in a wide frequency
band. Folding of the data on the frequencies with $\sim 3\,\sigma$
significance uncorrected for bandwidth demonstrates that none yields a
sinusoidal modulation with an amplitude exceeding either 0.004\,mag in
\ir\ or 0.4\% in \vr.  Since the addition of an artificial sinusoidal
signal to the data produces a peak in the periodogram with FWHM $\sim$
3 pixels, the number of independent trial periods within the total
bandwith is reduced to $\sim 2.300$. Corrected for the bandwith, the
Fisher-Snedecor critical value for a 3-$\sigma$ detection of a
periodic signal in the data then is $\Theta_{\rm cr} = 4.39$.  We
conclude that there is no evidence for periodic variation in both the
photometric and the polarimetric variation.

\begin{table}[t]
\caption[ ]{\label{poltab} Circular polarization in the red part of
the spectrum obatined from spectropolarimetry and polarimetry in the
Bessell R--band and in quasi R--bands. $\lambda_{\rm c}$ and $\Delta
\lambda$ refer to the central wavelength and the width of the
respective band. }
\begin{flushleft}
\begin{tabular*}{\hsize}{@{\extracolsep{\fill}}lccccl}
\noalign{\smallskip} \hline \noalign{\smallskip}\\
Date & HJD&$\lambda_{\rm c}$&$\Delta \lambda$&$V_{\rm R}$& \hspace{-2mm}Ref.\\
     &    & (\AA)           & (\AA)          & (\%)      &    \\[0.3ex] 
\noalign{\smallskip} \hline \noalign{\smallskip}
1977 Feb 22   & 2443196.5 & 6250 & 1500 & \hspace{2mm}$ 10.2 \pm 0.5$ & 1 \\
\hspace{8mm}Feb 23 & 2443197.5 & & & \hspace{2mm}$ 10.4 \pm 0.5$ & 1 \\
1986          &           & 6500 & 1300 & $10.0 \pm 0.7$ & 2 \\
1987 Feb 23.9 & 2446850.4 & 7000 & 1000 & \hspace{2mm}$8.2 \pm 1.0$  & 3 \\
1994 May 7    & 2449480.4 & 6550 & 1700 & \hspace{2mm}$6.0\pm 0.5$   & 4 \\
2000 Feb 4    & 2451578.6 & 6550 & 1700 & \hspace{2mm}$9.1\pm 0.3$   & 5 \\
2000 Jul 3/4  & 2451729.5 & 6100 & \hspace{1.3mm}800 & \hspace{2mm}$9.5\pm 0.3$   & 6 \\
\noalign{\smallskip}\hline\noalign{\smallskip}
\end{tabular*}
{\small References: (1) Liebert et al. 1978, (2) West 1989, (3) Robert
\& Moffat 1989, (4) Schmidt et al. 1995, (5) This work, (6) Jordan \&
Friedrich 2001.}
\end{flushleft}
\vspace*{-3mm}
\end{table}

The mean of our $R-$band circular polarization values is $V_{\rm R} =
9.1 \pm 0.3$\%.  We summarize this and other published polarization
measurements in the red part of the spectrum (quasi R--bands) in
Tab. 1.  The spectropolarimetric observations of Liebert et al. (1978)
and Schmidt et al. (1995) have been averaged over pass bands as close
as possible to the Kron-Cousins $R$-band. Since the circular
polarization falls off from a maximum at 5750\,\AA\ towards longer
wavelengths (Schmidt et al. 1995, their Fig. 2), the values quoted in
Table 1 are not strictly comparable. We have tried to account for this
uncertainty in assigning the errors. The quoted circular polarizations
are consistent with each other, except for the low value measured from
the polarization spectrum of Schmidt et al. (1995). Taken at face
value, the data in Table 1 suggest that there is a variation over the
time span of 23 years with a period of this order. The sparcity of the
data calls, however, for more extensive polarimetric monitoring with a
single instrument, either broad band polarimetry or preferably
spectropolarimetry as described by Jordan \& Friedrich (2001)

\section{Discussion and Conclusion}

We have performed a sensitive search for short periodicities in the
$R$-band flux and circular polarization of the highly magnetic white
dwarf LP790-29 which was previously thought to have a rotational
period \prot\ $> 100$\,yrs (Schmidt \& Norsworthy 1991). We undertook
this search because the sparse data available so far may have
prevented the discovery of short periods.

Rapid rotation would be the signature of a white dwarf spun up in a
cataclysmic variable of the AM Herculis type (Meyer \&
Meyer-Hofmeister (1999), of the AM Canes Venaticorum type (Iben \&
Tutukov 1991), or in a double degenerate merger (e.g. Segretain et
al. 1997, Ferrario et al. 1997). 
A merger may rotate at the disruption limit.  Segretain et al. argued,
however, that the merger loses 90\% of the initial angular momentum by
a strong wind, yielding an initial rotational period \prot\,$\sim
1$\,min. At an effective temperature of about 8000\,K (Liebert et
al. 1978, Bues 1999) the cooling age of \lp\ is $t_{\rm cool} \simeq 2
\times 10^9$\,yrs (Anselowitz et al. 1999), but may be shorter if it
originated from the merger of a cool white dwarf with its companion
(Segretain et al. 1997). Hence, even a cool white dwarf might still be
a fast rotator, although over the time the original rotational
velocity may have been reduced by magnetic braking.



Our principal result is the absence of variability in \lp\ with
periods between 4\,sec and about 3 hours and amplitudes $\Delta R >
0.009$\,mag and $\Delta V_{\rm R} > 0.7$\,\%.  This includes the
absence of photometric variability of the type one might expect in a
short period binary. Hence, there is no positive evidence for fast
rotation and no evidence for any of the above scenarios.



The only remaining possibilities which could mask rapid rotation in
\lp\ are (i) the almost perfect alignment of the rotational and
magnetic axes of an \mbox{azimuthally} symmetric field or (ii) a
rotational axis oriented directly towards the observer, leading to
rotational variability below our detection limit of the circular
polarization.

Previous circular polarimetry excludes periods longer than
\mbox{\prot\ $\sim 1$\,h}, although the limit \prot\ $> 100$\,yrs set
by West (1989) and Schmidt \& Norsworthy (1991) may be premature in
view of the low level of the 1994 circular polarization by Schmidt et
al. (1995). Nevertheless, these results suggest that \lp\ is, in fact,
an exceedingly slow rotator. It seems worthwhile to follow up the
possibility of a period of about a quarter of a century (see also the
paper by Jordan \& Friedrich 2001) by monitoring the
level of circular polarization.
%

\acknowledgements{We thank Stefan Jordan, Boris T. G\"ansicke,
Frederic V. Hessman for valuable comments on the manuscript. This work
was supported in part by BMBF/DLR grant 50\,OR\,9903\,6.}

\end{document}